\documentstyle[12pt]{article}
\begin{document}
\begin{center}{\large{\bf Mapping Among Manifolds III}}
\end{center}
\vspace*{1.5cm}
\begin{center}
A. C. V. V. de Siqueira
$^{*}$ \\
Departamento de Educa\c{c}\~ao\\
Universidade Federal Rural de Pernambuco \\
52.171-900, Recife, PE, Brazil\\
\end{center}
\vspace*{1.5cm}
\begin{center}{\bf Abstract}

In two previous papers, we constructed two modified Hamiltonian formalisms
to make maps among manifolds explicit. In this paper, the two
modified formalisms were adapted to manifolds with local coordinates given by
scalar fields, as in a classical nonlinear sigma-model. The scalar field
coordinates could be built from vectors, tensors, spinors, Lagrangians, Hamiltonians, group parameters, etc.
\end{center}

 \vspace{3cm}

${}^*$ E-mail: acvvs@ded.ufrpe.br
\newline

\newpage

\section{Introduction}
$         $

In two previous papers, we constructed two modified Hamiltonian formalisms
to make maps among manifolds explicit \cite{1}, \cite{2}.
In this paper, the two modified formalisms were adapted to manifolds with local coordinates given by
scalar fields, as in the classical nonlinear sigma-model \cite{3}. The scalar
field coordinates could be built from vectors, tensors, spinors, Lagrangians, Hamiltonians, group parameters, etc.
This paper is organized as follows: In Sec.$2$, we build the third modified Hamiltonian formalism by exchanging the
set of ordinary local coordinates $ (x^{\mu})$ used in the first and second
modified Hamiltonian formalisms for a set of scalar fields $Q^{(\mathbf{A})}$, taking
the first step toward the third modified Hamiltonian formalism; in Sec.$3$ (Concluding Remarks),
we present differences between the third modified Hamiltonian formalism and the other formalisms.
\newpage
\renewcommand{\theequation}{\thesection.\arabic{equation}}
\section{\bf The Modified Formalism III}
\setcounter{equation}{0}
 $         $
Before giving a brief overview of the third modified Hamiltonian
formalism, we make some important observations. Notice that we exchange
the set of ordinary local coordinates $ (x^{\mu})$ in an ordinary manifold
for the set of scalar fields $Q^{(\mathbf{A})}$, which are functions of some
ordinary local coordinates $ (x^{\mu})$, as in the classical nonlinear sigma-model.
It is sometimes possible to define a line element in the manifold. In this case, let us
consider the line element as
\begin{equation}
 ds^2= G_{\Lambda\Pi}(Q)dQ^{\Lambda}dQ^{\Pi},
\end{equation}
with
\begin{equation}
G_{\Lambda\Pi}(Q)=E_{\Lambda}^{(\mathbf{A})}(Q)E_{\Pi}^{(\mathbf{B})}(Q)\eta_{(\mathbf{A})(\mathbf{B})},
\end{equation}
in which $ \eta_{(\mathbf{A})(\mathbf{B})}$ and $
E_{\Lambda}^{(\mathbf{A})}(Q)$ are flat metric and vielbein
components, respectively. Notice that the set of scalar fields $Q^{(\mathbf{A})}$
is given by $(Q^{\Pi}E_{\Pi}^{(\mathbf{A})})$ and $ (Q^{\Pi})$ are the local coordinates in the
new manifold. We could build a set of scalar fields from vectors, tensors, spinors, etc. We have
also well-defined curves in the manifold as a function of local coordinates $ (Q^{\Pi})$ with an
evolution parameter $ t $.
\newline
We choose each $ \eta_{(\mathbf{A})(\mathbf{B})}$ as plus or minus Kronecker's delta function.
\newline
The third modified Hamiltonian formalism is the second modified Hamiltonian formalism with scalar fields as variables of the
configuration space. In this case, we have the usual properties employed in the second modified
Hamiltonian formalism.
In a more general case, as well as that given by (2.1) and (2.2), we consider an $H({t})$ as a $t$ parameter-dependent function.
Let us define 2n variables that will be called ${\xi}^j$ with index j running from 1 to 2n so that we have
\newline
${\xi}^j$ $\in$ $({\xi}^1,\ldots,{\xi}^n,{\xi}^{n+1},\ldots,{\xi}^{2n})$ =$({Q}^1,\ldots,{Q}^n,{P}^1,\ldots,{P}^n)$
in which ${Q}^j$ and ${P}^j$ may or may not be the usual coordinates and momenta, respectively, in the new manifold.
We now define the function by
\begin{equation}
 H({\tau})=\frac{1}{2}H_{ij}{\xi}^i{\xi}^j,
\end{equation}
in which $H_{ij}$ is a symmetric matrix. Consider the following
system:
\begin{equation}
\frac{d{\xi}^i}{d\tau
}=I_{1}^{ik}\frac{\partial{H}}{\partial{\xi}^k } .
\end{equation}
Equation (2.4) introduces the $I_{1}$ given by
\begin{equation}
\left(%
\begin{array}{cc}
  O & A \\
  B & O \\
\end{array}%
\right)
\end{equation}
in which O, A and B are the $n \textbf{x}n$. O is the zero
matrix. $A=\epsilon_{1}I$ and $B=\epsilon_{2}I$ are
proportional to the identity matrix, with $\epsilon_{i}=-1,+1$ and
$i=1$ or 2. We now make a linear transformation from ${\xi}^j$ to
${\eta}^j$ given by
\begin{equation}
  {\eta}^j={{T}^j}_k{\xi}^k,
\end{equation}
in which ${{T}^j}_k$ is a non-symplectic matrix and the new function
is given by
\begin{equation}
 \bar{H}=\frac{1}{2}C_{ij}{\eta}^i{\eta}^j,
\end{equation}
in which $C_{ij}$ is a symmetric matrix. Consider that (2.7) obeys
the following equation
\begin{equation}
\frac{d{\eta}^i}{d\tau
}=I_{2}^{ik}\frac{\partial{H}}{\partial{\eta}^k },
\end{equation}
in which $I_{2}$ is given by
\begin{equation}
\left(%
\begin{array}{cc}
  O & E \\
  D & O \\
\end{array}%
\right)
\end{equation}
and O, E and D are $n \textbf{x}n$. O is the zero matrix.
$E=\epsilon_{3}I$ and $D=\epsilon_{4}I$ are proportional to
the identity matrix, with $\epsilon_{j}=-1,+1$ and $j=3$ or 4.
The functions A, B, E and D could be chosen as arbitrary diagonal
matrices. However, such a possibility will not be used in this
paper. The matrices H, C and T obey the following system
\begin{equation}
 \frac{d{{T}^i}_j}{d\tau}+\frac{d{t}}{d\tau}{{T}^i}_kI_{1}^{kl}X_{lj}=I_{2}^{im}Y_{ml}{{T}^j}_k,
\end{equation}
in which $2X_{lj}=\frac{\partial{H_{ij}}}{\partial{\xi}^l
}\xi^{i}+2H_{lj}$ and
$2Y_{ml}=\frac{\partial{C_{il}}}{\partial{\eta}^m
}\eta^{i}+2C_{ml}.$ t and $\tau$ are the evolution parameters of two curves in two different manifolds.
We note that (2.10) is a first-order linear differential equation system in ${{T}^i}_k $ and
that the non-linearities in the Hamiltonians were transferred to
their coefficients. Consider $\frac{d{t}}{d\tau}X_{lj}=Z_{lj}$ and
write (2.10) in the matrix form
\begin{equation}
 \frac{d{T}}{d\tau}+TI_{1}Z=I_{2}YT,
\end{equation}
in which T, Z and Y are $2n \textbf{x}2n$ matrices as
\begin{equation}
\left(%
\begin{array}{cc}
  T_{1} & T_{2} \\
  T_{3} & T_{4} \\
\end{array}%
\right)
\end{equation}
with similar expressions for Z and Y. Let us write (2.11) as
follows:
\begin{equation}
 \dot{T_1}=\epsilon_{3}(Y_{3}T_{1}+Y_{4}T_{3})-\epsilon_{2}T_{2}Z_{1}-\epsilon_{1}T_{1}Z_{3},
\end{equation}
\begin{equation}
 \dot{T_2}=\epsilon_{3}(Y_{3}T_{2}+Y_{4}T_{4})-\epsilon_{2}T_{2}Z_{2}-\epsilon_{1}T_{1}Z_{4},
\end{equation}
\begin{equation}
 \dot{T_3}=\epsilon_{4}(Y_{1}T_{1}+Y_{2}T_{3})-\epsilon_{2}T_{4}Z_{1}-\epsilon_{1}T_{3}Z_{3},
\end{equation}
\begin{equation}
 \dot{T_4}=\epsilon_{4}(Y_{1}T_{2}+Y_{2}T_{4})-\epsilon_{2}T_{4}Z_{2}-\epsilon_{1}T_{3}Z_{4}.
\end{equation}
Now consider
\begin{equation}
 \dot{S_1}=\epsilon_{3}(Y_{3}S_{1}+Y_{4}S_{3}),
\end{equation}
\begin{equation}
 \dot{S_2}=\epsilon_{3}(Y_{3}S_{2}+Y_{4}S_{4}),
\end{equation}
\begin{equation}
 \dot{S_3}=\epsilon_{4}(Y_{1}S_{1}+Y_{2}S_{3}),
\end{equation}
\begin{equation}
 \dot{S_4}=\epsilon_{4}(Y_{1}S_{2}+Y_{2}S_{4}),
\end{equation}
and
\begin{equation}
 \dot{R_1}=-\epsilon_{2}R_{2}Z_{1}-\epsilon_{1}R_{1}Z_{3},
\end{equation}
\begin{equation}
 \dot{R_2}=-\epsilon_{2}R_{2}Z_{2}-\epsilon_{1}R_{1}Z_{4},
\end{equation}
\begin{equation}
 \dot{R_3}=-\epsilon_{2}R_{4}Z_{1}-\epsilon_{1}R_{3}Z_{3},
\end{equation}
\begin{equation}
 \dot{R_4}=-\epsilon_{2}R_{4}Z_{2}-\epsilon_{1}R_{3}Z_{4}.
\end{equation}
From the theory of first-order differential equation systems,
it is well known that each system in (2.17)-(2.24) has a solution in the
region where $Z_{lj}$ and $Y_{ml}$ are continuous functions. In
this case, the solution for (2.10) or (2.11) is given by
\begin{equation}
 {T_1}=(S_{1}a+S_{2}b)R_{1}+(S_{1}d+S_{2}c)R_{3},
\end{equation}
\begin{equation}
{T_2}=(S_{1}a+S_{2}b)R_{2}+(S_{1}d+S_{2}c)R_{4},
\end{equation}
\begin{equation}
{T_3}=(S_{3}a+S_{4}b)R_{1}+(S_{3}d+S_{4}c)R_{3},
\end{equation}
\begin{equation}
{T_4}=(S_{3}a+S_{4}b)R_{2}+(S_{3}d+S_{4}c)R_{4},
\end{equation}
\newpage
in which a, b, c and d are constant $n \textbf{x}n$ matrices and, by placing
(2.25)-(2.28) into (2.6), we will have completed the mapping among
manifolds.
\newline
By exchanging the scalar field coordinates for the ordinary local
coordinates, we recover the second formalism. The second formalism can be
reduced to the first by an appropriate choice of the constants
$\epsilon_{i}$ and $\epsilon_{j}$.
\newline
Notice that, in general, the parameters t and $\tau$ are
Lie's parameter symmetries of the associated linear partial
differential systems obtained from the scalar field coordinates.
For example, we recall the nonlinear sigma-model, \cite{3}, in which
the symmetries are different from Lie's parameter symmetries above.
\newpage
\section{Concluding Remarks}
  $              $
It is very important to notice that the Hamiltonians that appear in this
paper are different from those employed in traditional field theory,
in which variational functional derivatives and partial derivatives in
relation to ordinary coordinates are used. It is also different from
the nonlinear sigma-model. In this paper, operators such as
$ \frac{\partial}{\partial(x^{\mu})}$ do not explicitly appear in the mapping. However, as we could have a set of manifolds in which
the scalar field coordinates could be a composition of a set of
ordinary Hamiltonians or Lagrangians, etc., the derivative dependence could be implicit.

\end{document}